\documentclass[fleqn,usenatbib]{mnras}

\usepackage{newtxtext,newtxmath}
\usepackage[T1]{fontenc}
\DeclareRobustCommand{\VAN}[3]{#2}
\let\VANthebibliography\thebibliography
\def\thebibliography{\DeclareRobustCommand{\VAN}[3]{##3}\VANthebibliography}

\usepackage{graphicx}	% Including figure files
\usepackage{amsmath}	% Advanced maths commands
\usepackage{enumerate}

\usepackage{multicol}        % Multi-column entries in tables
\usepackage{pdflscape}	% Landscape pages

\usepackage{ulem}

\usepackage[T1]{fontenc}
\usepackage{ae,aecompl}

\usepackage{newtxtext,newtxmath}
\newcommand\kms{\mbox{\,km s}^{-1}}

\def\promedio#1{\langle{#1}\rangle}

\def\apss{Ap\& SS}
\def\apj{ApJ}
\def\apjl{ApJL}
\def\aap{A\& A}

\def\mnras{MNRAS}

\def\apjs{ApJS}

\def\mnras{{MNRAS}}
\def\pasj{{PASJ}}

\title[Core separation in turbulent MCs]{The sonic scale does not determine the core separation in turbulent molecular clouds}

\author[Zamora-Aviles et al.]{
Manuel Zamora-Aviles$^{1}$,
Javier Ballesteros-Paredes$^2$\thanks{Contact e-mail: \href{j.ballesteros@irya.unam.mx}{j.ballesteros@irya.unam.mx}},
Aina Palau$^{2}$,
 Enrique V\'azquez-Semadeni$^{2}$,
\and Gilberto C. G\'omez$^{2}$
\\
$^{1}$Instituto Nacional de Astrof\'isica, \'Optica y Electr\'onica, Luis E. Erro 1, 72840 Tonantzintla, Puebla, M\'exico\\
$^{2}$Universidad Nacional Autónoma de México, Instituto de Radioastronom\'ia y Astrof\'isica, Campus Morelia. Apartado Postal 3-72, 58089 Morelia Michoac\'an, M\'exico\\
}

\date{Accepted XXX. Received YYY; in original form ZZZ}

\pubyear{2015}

\begin{document}
\label{firstpage}
\pagerange{\pageref{firstpage}--\pageref{lastpage}}
\maketitle

% Abstract of the paper
%It should be a single paragraph not more than 250 words (200 words for Letters).
\begin{abstract}

It has recently been suggested that the typical separation between cores in molecular clouds dominated by turbulence is determined by the sonic scale, the size scale at which the turbulent velocity dispersion equals the sound speed. In this work, we test this hypothesis using a suite of turbulent simulations with Mach numbers $\mathcal{M}=4$ and 8, and three turbulent forcing wavenumbers ($k_{\rm for}=2,4$ and 8). Dense cores are identified through dendrogram analysis of column density maps, and their separations are compared to the sonic scale measured from velocity structure functions. We find no statistical correlation between the core separation and the sonic scale nor with the driving scale. Instead, for each run, the core separation spans the entire range of values between these two scales. Our results indicate that fragmentation in turbulence-dominated clouds is not governed by an intrinsic scale in the turbulent cascade. This finding calls into question the use of the sonic scale as a predictive quantity in star formation theories and cautions against interpreting observational core spacings as evidence for universal turbulent fragmentation physics.

\end{abstract}

% Select between one and six entries from the list of approved keywords.
% Don't make up new ones.

%%%%%%%%%%%%%%%%%%%%%%%%%%%%%%%%%%%%%%%%%%%%%%%%%%

%%%%%%%%%%%%%%%%% BODY OF PAPER %%%%%%%%%%%%%%%%%%
\defcitealias{Ishihara2025}{I25}
\section{Introduction} \label{sec:introduction}

Star formation %{\steal{is initiated} 
occurs %[EVS: El momento del ``inicio'' no est\'a claramente definido.]} 
within dense cores embedded in molecular clouds (MCs), yet the physical mechanisms responsible for the formation and spatial distribution of these cores remain an open question. Molecular cloud fragmentation can be influenced by the interplay of various processes, including thermal pressure, supersonic turbulence, magnetic fields, large-scale flows, and gravitational %{\steal{instabilities \textcolor{red}{such as the Jeans}} 
instability. A central issue in star formation theory is identifying which of these mechanisms dominates in setting the characteristic properties of dense cores, such as their masses and separations. This has been widely studied in gravity-dominated dense clouds and clumps, where an increasing number of works point to a dominant role of thermal Jeans processes regulating fragmentation \citep[e.g.,][]{Gutermuth2011, Palau2015, Beuther2018, Li2019, Lin2019, Palau2021, Sanhueza2019, Svoboda2019, Morii2024, Xu2024, Das2024}.
 
In contrast, in low-mass, low-density, and apparently turbulence-dominated environments such as the Polaris Flare and Lupus I clouds, it has been recently suggested that many cores seem to be gravitationally unbound and exhibit separations much smaller than the Jeans length \citep{Ishihara2025}, %{\steal{. \textcolor{red}{Thus, the properties of the cores observed in this kind of clouds, potentially affected by strong stellar feedback,} challenge} , 
thus challenging the applicability of pure gravitational fragmentation in such clouds.

An alternative framework to describe fragmentation in these low-density, apparently turbulence-dominated regions is provided by turbulent fragmentation models, in which supersonic turbulence generates a network of shocks that naturally leads to density enhancements \citep[e.g.,][]{Vazquez-Semadeni+1994, Passot_VS98, Padoan+01, Padoan+20, Padoan_Nordlund2002, VS+03, Ballesteros-Paredes04b, MacLow_Klessen2004, Ballesteros-Paredes+06, Ballesteros-Paredes+07, Henneb_Chabrier08, Hopkins12, Federrath_Klessen2012}. In this scenario, the density { fluctuation} field { under isothermal conditions} develops a log-normal probability distribution function (PDF), and fragmentation is expected to occur { at all scales between the turbulence injection scale and \textit{the sonic scale}, the spatial scale where the turbulent velocity dispersion equals the sound speed \citep[e.g.,][]{VS+03}}. Recently, \citeauthor{Ishihara2025} (\citeyear{Ishihara2025}; hereafter, \citetalias{Ishihara2025}) suggested that the observed core separations in { the} Polaris Flare \citep[$0.27$~pc;][]{Ossenkopf_MacLow2002} and Lupus I \citep[$0.13$~pc; ][]{Brunt2010,Yun+2021a,Yun+2021b} { clouds} closely match the estimated sonic scales, { and thus proposed} that turbulence, rather than gravity, regulates core formation in these environments, { so that the core separation corresponds to the sonic length of the turbulence instead}. 

{ However, this suggestion is questionable because of  mainly two reasons. {\it a}) In their Sec. 4, \citetalias{Ishihara2025} argue that the typical core separation is 20-40 times smaller than the {\it global} (i.e., cloud-scale) Jeans length. However, this does not appear to be the relevant Jeans length to compare to, since there are several levels of the {\sc Dendrograms} hierarchy (cloud, trunk, branch, and leaf) between the cloud-scale and the core scale. The relevant Jeans length is that of the immediate parent structures of the cores. Since the cores correspond to the``leaves'', the relevant Jeans length should be that of the ``branches'', not that of the whole clouds. Those authors further compare the core masses with the ``local'' Jeans masses $m_{\rm J}$, but they do not specify for which level of the dendrogram hierarchy this local Jeans mass is measured. {\it b})  \citetalias{Ishihara2025} do not explain their hypothesis that the sonic scale should be represented by the {\it separation} between the cores, rather than by their sizes.}

{ In any case}, the sonic scale hypothesis remains largely untested in controlled numerical experiments. In particular, it is unclear whether fragmentation in turbulent clouds naturally produces core separations that correlate with the sonic scale, or whether other parameters, such as the turbulent forcing scale, play a more dominant role. Although the sonic scale { does} reflect a local transition in the turbulent cascade, the large-scale properties of the turbulence may strongly influence fragmentation. In this work, we { thus} address this question through a set of turbulent numerical simulations. We systematically vary the Mach number and the turbulent forcing scale and analyse the resulting distribution of dense cores using dendrogram-based structure identification techniques \citep{Rosolowsky2008}. { We} focus on quantifying the characteristic core separation and assessing its dependence on the sonic scale and the energy injection scale. By comparing our simulation results with observational findings in nearby clouds, particularly those presented by \citetalias{Ishihara2025}, we aim to evaluate the predictive power of the sonic-scale hypothesis and identify the dominant physical processes responsible for fragmentation in low-mass molecular clouds. This has important implications for interpreting core statistics in both nearby and distant star-forming regions and for understanding the physical origins of the initial conditions of star formation.

\section{Numerical methods} \label{sec:numerics}

\subsection{Turbulent simulations}

We perform a suite of 3D numerical simulations of { supersonic isothermal turbulence} using the adaptive mesh refinement code \textsc{FLASH} \citep[V4.7.1;][]{FLASH1,FLASH2,FLASH3}. The simulations are carried out in a periodic, cubic domain. Although the simulations are isothermal and thus reescalable, we provide a set of physical units to allow the reader to compare with actual clouds. Hence, adopting a constant temperature of $T=15$~K, the isothermal sound speed is $c_{\rm s}~\approx~0.24~\kms$, the length of the box results in $10~\mathrm{pc}$, the total mass in $5 \times 10^3~M_\odot$, and the mean density in $\langle \rho \rangle \approx 3.4 \times 10^{-22}~\mathrm{g~cm^{-3}}$ (or $\langle n \rangle \approx 88 ~\mathrm{cm^{-3}}$ assuming a mean molecular weight of 2.3). 

All models were executed with a fixed resolution of $512^3$ grid cells (or $\Delta x \approx 0.019$ pc). Turbulence is driven by a stochastic forcing scheme in Fourier space as described and implemented in the FLASH code by \citet{Federrath+2009}. We carried out two sets of simulations in which kinetic energy was continuously injected to maintain a constant Mach number, set to $\mathcal{M} = 4$ and $\mathcal{M} = 8$, respectively. Within each set, three runs were performed, differing by the spatial scale of energy injection: large-scale driving with wavenumbers in the range $1 \leq k \leq 2$, intermediate-scale driving with $2 \leq k \leq 4$, and small-scale driving with $4 \leq k \leq 8$. The forcing is a natural mixture of solenoidal and compressive modes in all runs \citep[{ with the compressibility parameter $b=0.5$;} see, e.g., ][]{Federrath+10b}. We evolve the simulations during four turbulent crossing times, $t_{\rm c}$. The analysis was performed at times $t=$~2, 3 and 4~$t_{\rm c}$ since statistically, after $\sim1-2$~$t_c$, the physical properties of each simulation do not change over time. For a summary of the numerical parameters, see Table \ref{tab:simulations}. 
%, corresponding to {41.3 and 20.6 Myr for the simulations with $\mathcal{M} = 4$ and $\mathcal{M} = 8$, respectively. 

To isolate the role of turbulence and eliminate the potential effects of gravitational { contraction}, we { have} deliberately excluded self-gravity from our simulations. While real MCs (even those of relatively low density such as Polaris) are influenced to some extent by self-gravity, the objective of this study is to test the hypothesis proposed by \citetalias{Ishihara2025}, { that turbulence alone determines the typical separation between cores, which is given by the sonic scale}.

\begin{table}
\centering
\caption{Summary of the numerical simulation models.
Each simulation is characterized by two parameters:
(1) the rms Mach number $\mathcal{M}$
and (2) the turbulent forcing scale, defined by the range of wavenumbers $k_{\rm for}$ over which energy is injected.
}
\begin{tabular}{lcc}
\hline\hline
Simulation & Mach Number ($\mathcal{M}$) & Forcing scale ($k_{\rm for}$ range) \\ %& Resolution \\
\hline
M4\_k1-2 & 4 & $1 \leq k_{\rm for} \leq 2$ \\ %& $1024^3$ \\
M4\_k2-4 & 4 & $2 \leq k_{\rm for} \leq 4$ \\ %& $1024^3$ \\
M4\_k4-8 & 4 & $4 \leq k_{\rm for} \leq 8$ \\ %& $1024^3$ \\
M8\_k1-2 & 8 & $1 \leq k_{\rm for} \leq 2$ \\ %& $1024^3$ \\
M8\_k2-4 & 8 & $2 \leq k_{\rm for} \leq 4$ \\ %& $1024^3$ \\
M8\_k4-8 & 8 & $4 \leq k_{\rm for} \leq 8$ \\ %& $1024^3$ \\
\hline
\end{tabular}
\label{tab:simulations}
\end{table}

% /mnt/e/myProjects/Turb/python/dendro/proj_M4_v2.py
\begin{figure*}
    \centering
\includegraphics[width=\textwidth]{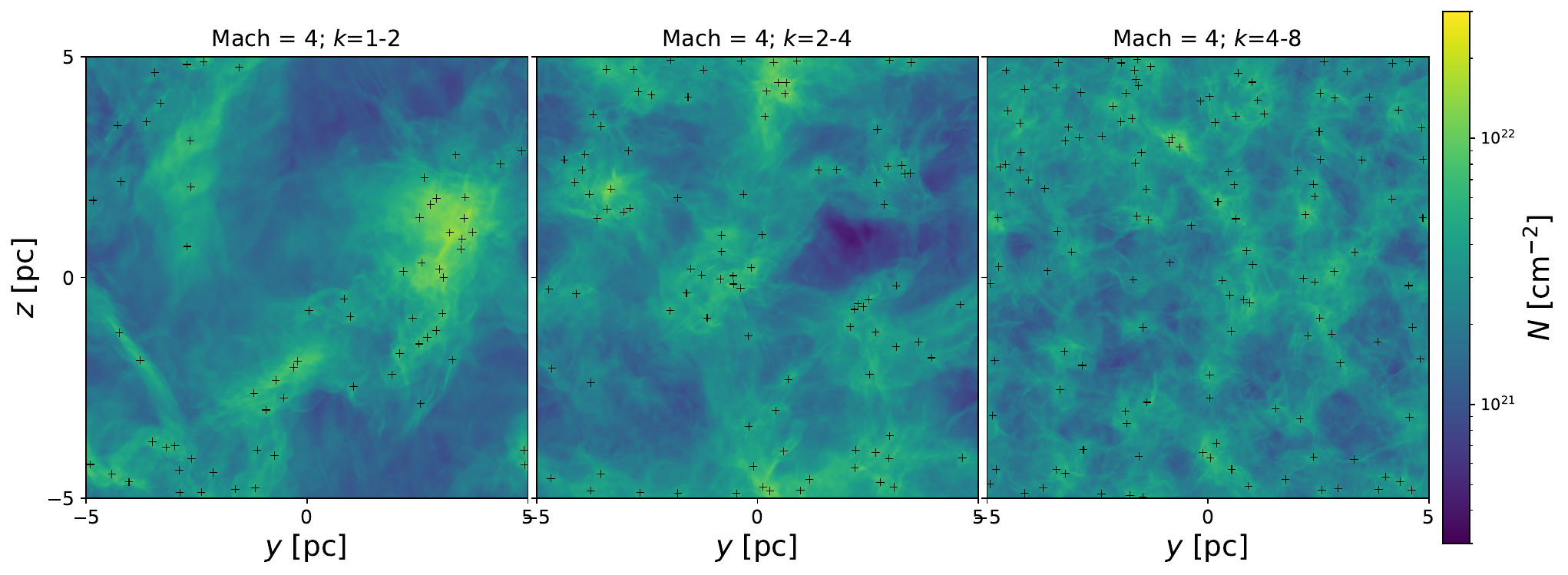}
    \caption{Column density maps { at $t=4 \times t_{\rm c}$} for Mach=4 models along the x–axis for forcing scales $k_{\rm for}=1$–$2$ (left), $k_{\rm for}=2$–$4$ (centre), and $k_{\rm for}=4$–$8$ (right).  Each panel displays the same logarithmic colour scale, with black plus symbols marking core centroids identified by the dendrogram.}
    \label{fig:M4}
\end{figure*}

%/mnt/e/myProjects/Turb/python/dendro/proj_M8_v2.py
\begin{figure*}
    \centering   \includegraphics[width=\textwidth]{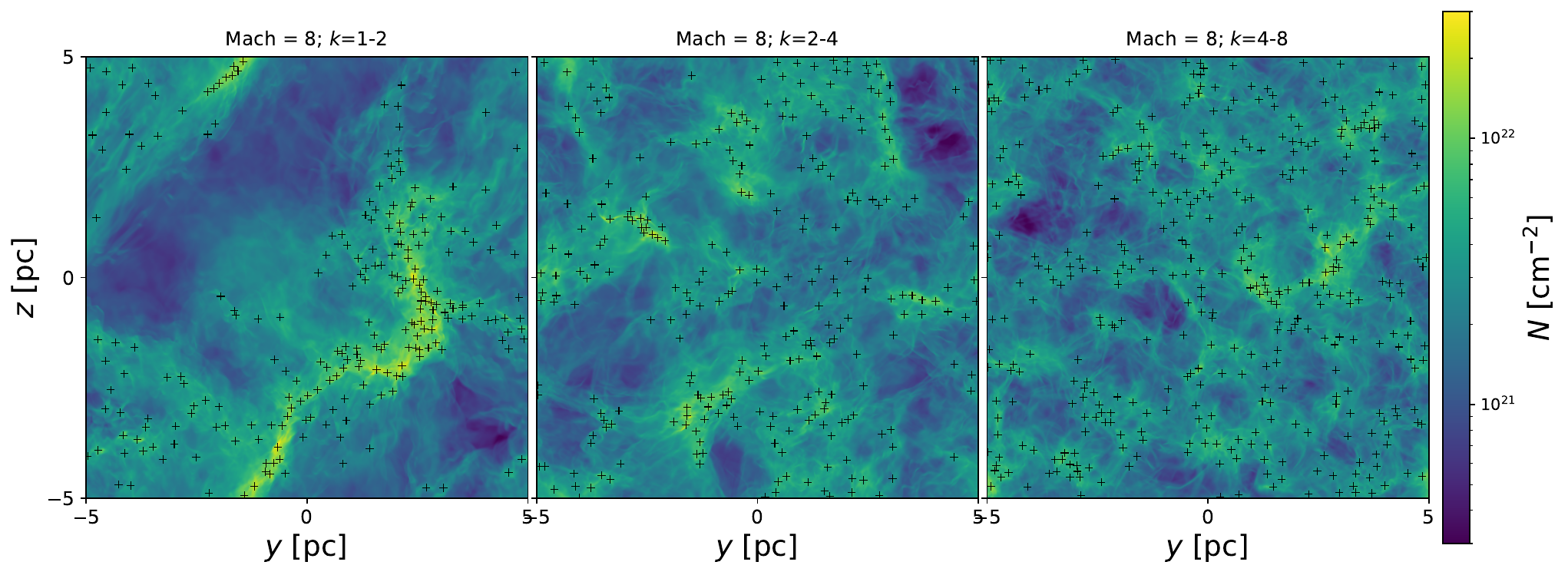}
    \caption{Same as Fig \ref{fig:M4} but for models with Mach=8.}
    \label{fig:M8}
\end{figure*}

%%%%%%%%%%%%%%%%%%
\subsection{Core identification with dendrograms}

In order to determine whether the sonic scale defines the structure of the density field and whether it has a relation to the core separation, we need to identify cores. For this purpose, we employed the \texttt{astrodendro} package \citep{Rosolowsky2008}, which constructs a hierarchical tree of all contiguous overdense regions above a specified baseline threshold, applied to the three main column density projections of each simulation.\footnote{{ Note that here we aim at simulating the observational procedure, and therefore define the cores in the two-dimensional projected data. Nevertheless, a more fundamental study would be to define the cores directly in the three-dimensional density data cube, and correlate their separations to the sonic scale, also computed from the 3D data. We defer this more detailed study to a future contribution.}} The algorithm requires two key input parameters, $\mathtt{min\_value}$, which defines the minimum contour level, and $\mathtt{min\_delta}$, which specifies the minimum contrast between adjacent nested structures, ensuring that only statistically and physically meaningful peaks are retained.\footnote{Detailed documentation for \texttt{astrodendro} is available online at \url{http://dendrograms.org}.}

Following \citet{Chen+2018}, we express the dendrogram thresholds relative to the median column density value of the map, $N_{0}$, adopting $\mathtt{min\_value} = 0.95~N_{0}$, $\mathtt{min\_delta} = 0.475~N_{0}$, and requiring each identified structure to contain at least $\mathtt{min\_npix}=9$ pixels, equivalent to a projected area of $(0.05~\mathrm{pc})^2$. As shown in Table~\ref{tab:dendro_params} (Appendix~\ref{app:astrodendro}), varying the \texttt{astrodendro} parameters affects both the number of cores and, to a lesser extent, the median core separation. Nevertheless, the parameters that we used are consistent with those used by \citet{Chen+2018}.

Within the dendrogram, the non-branching elements, or ``leaves'', representing the smallest self-contained overdense regions, are extracted. We further applied an additional core-selection criterion following \citetalias{Ishihara2025}, after the approach of \citet{Takemura+2021}, in which only leaves with peak column densities exceeding twice the $\mathtt{min\_value}$ threshold are retained and classified as dense cores. From this point on, we refer to these dendrogram leaves simply as ``cores''.\footnote{{ Note that, although this procedure mimics the approach frequently used in observations, it has the problem that not all ``cores'' have the same physical properties, since the leaves extracted in lower-column density regions will likely be larger and less dense than those extracted in higher-density regions.}}

%%%%%%%%%%%%%%%%%%%%%%%%%
\subsection{Core separation and sonic scale} \label{subsec:cs}

 Following \citetalias{Ishihara2025}, the characteristic two-dimensional core separation ($S_{\rm 2D}$) was determined using the nearest-neighbour search algorithm (NNS) \citep{Knuth1973}, applied to all identified cores in each column density map. 
 This statistic provides a quantitative measure of the typical spacing between neighbouring cores, which we used to examine systematic variations across different turbulent Mach numbers and energy-injection scales.

{ On the other hand, w}hereas in the observational study by \citetalias{Ishihara2025} the sonic scale is assumed to be approximately $\ell_s\sim$~0.1~pc, in this work we determine it self-consistently by computing the second‐order velocity structure function,
\begin{equation}
  \mathcal{S}_{2}(\ell) = \bigl\langle \bigl\| \mathbf{v}(\mathbf{x} + \boldsymbol{\ell}) - \mathbf{v}(\mathbf{x})\bigr\|^{2} \bigr\rangle_{\mathbf{x}}~,  
\end{equation}
where { $\boldsymbol{\ell}$ is the spatial {\it lag}, and} the average is taken over all spatial positions $\mathbf{x}$ and along the three dimensions. The corresponding characteristic velocity difference is then defined as $\delta v(\ell) \;=\; \sqrt{\mathcal{S}_{2}(\ell)}$, and the sonic scale, $\ell_s$, is identified as the spatial scale at which $\delta v(\ell)$ equals the sound speed, marking the transition from supersonic to subsonic turbulent motions.

\section{Results}\label{sec:results}

Figures~\ref{fig:M4} and \ref{fig:M8} show column density maps projected along the $x$-axis for $\mathcal{M} = 4$ and 8 simulations, respectively, with black crosses marking the locations of the identified cores. From left to right, the panels correspond to turbulent forcing wavenumbers ranges of $k_{\rm for} = 1$–2, 2–4, and 4–8. These maps illustrate how the spatial distribution and clustering of cores vary with the Mach number and forcing scale.

% /part1/mzamora/Jupyter/Turb/plot_structure_function.ipynb
\begin{figure}
    \centering    \includegraphics[width=0.49\textwidth]{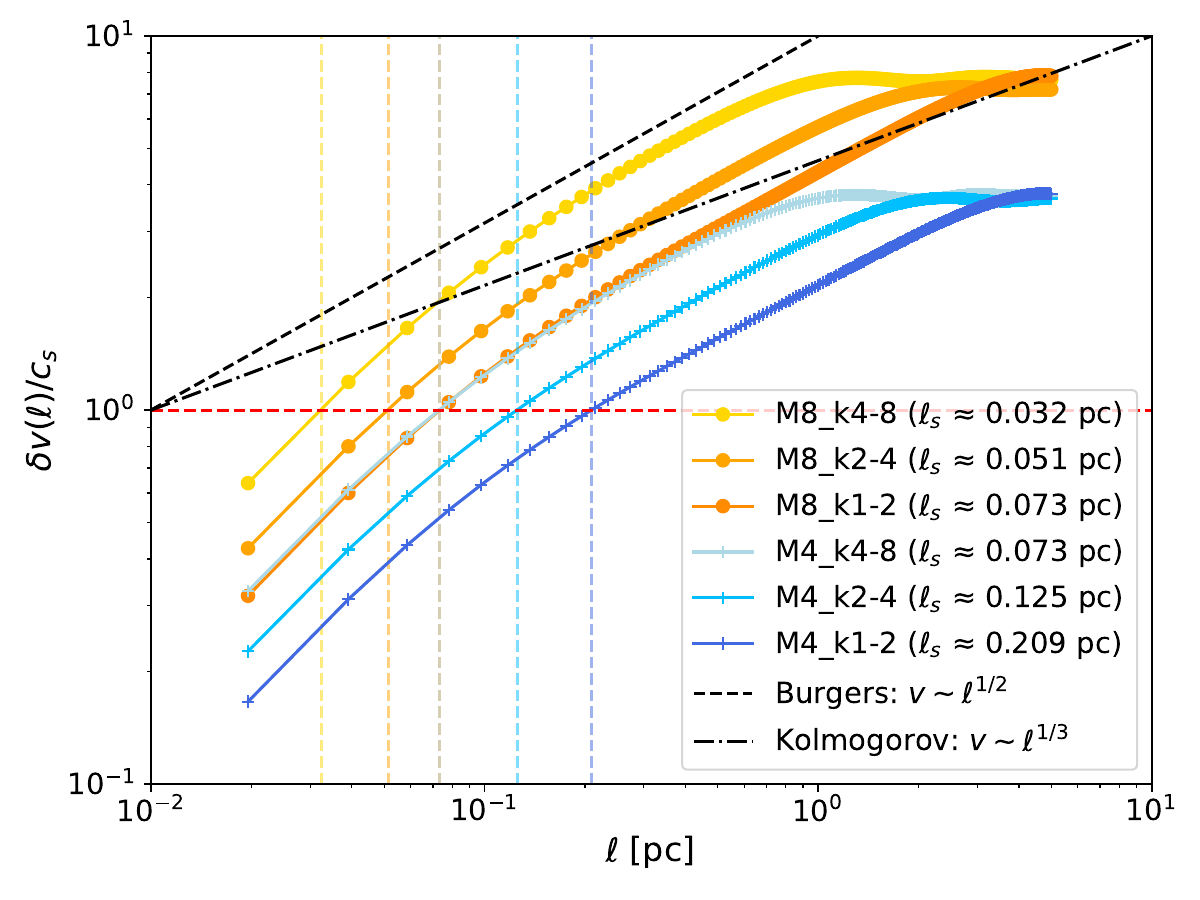}
    \caption{Characteristic velocity differences $\delta v(\ell)$, normalized by the sound speed $c_s$ as a function of spatial { lag} $\ell$, for all models. The horizontal red dashed line denotes the sonic scale, defined by the condition $\delta v(\ell_{s}) / c_s = 1$, while the corresponding value of $\ell_s$ for each model is indicated in the legend { and denoted by the vertical dashed lines}. Each curve represents a distinct combination of turbulent Mach number and forcing scale, as specified in the labels.}
    \label{Fig:sonic_scale}
\end{figure}

%2D 
% /mnt/e/myProjects/Turb/python/plot_histo_M4_v5.py plot_histo_M8_v5.py
% Statistics at t = 3 and 4 * t_c,turb, combining the three main projections:
\begin{table*}
  \centering
  \caption{Summary of the core statistics for all simulations in projection (2D). 
    The columns list: 
    (1) the simulation label, including Mach number (M4 or M8), forcing scale ($k$-range);
    (2) the physical forcing scale, $\lambda_{\rm for} = L_{\rm box}/k_{\rm for}$;
    (3) the sonic scale, $\ell_s$; 
    (4) total number of identified cores, $N_{\rm cores}$; 
    (5) the mean projected nearest-neighbor separation, $\langle S_{\rm 2D}\rangle$; 
    (6) the median projected separation, $S_{\rm 2D,med}$;
    (7) the standard deviation of the projected separation, $\sigma_{\rm 2D}$; 
    (8) the ratio $S_{\rm 2D,med}/\lambda_{\rm for}$; 
    (9) the ratio $S_{\rm 2D,med}/\ell_s$.}
  \label{tab:core_stats}
  \begin{tabular}{lcccccccc}
    \hline
    Model & $\lambda_{\rm for}$ [pc] & $\ell_s$ [pc] & $N_{\rm cores}$  & $\langle S_{\rm 2D}\rangle$ [pc] & $S_{\rm 2D,med}$ [pc] & $\sigma_{\rm 2D}$ [pc] & $S_{\rm 2D,med}/\lambda_{\rm for}$ & $S_{\rm 2D,med}/\ell_s$ \\
    \hline
%   Model       l     ls       N      <S>     Smed   sigma  Smed/l Smed/ls     
    M4\_k1-2 & 5.0 & 0.209 &  414 & 0.587 & 0.468 & 0.398 & 0.094 & 2.24 \\
    M4\_k2-4 & 2.5 & 0.125 &  635 & 0.491 & 0.417 & 0.312 & 0.167 & 3.34 \\
    M4\_k4-8 & 1.25 & 0.073 & 716 & 0.552 & 0.486 & 0.311 & 0.389 & 6.66 \\
    \hline
    M8\_k1-2 & 5.0 & 0.073 &  1813 & 0.273 & 0.230 & 0.162 & 0.046 & 3.15 \\
    M8\_k2-4 & 2.5 & 0.051 &  2195 & 0.276 & 0.237 & 0.155 & 0.095 & 4.65 \\
    M8\_k4-8 & 1.25 & 0.032 & 2579 & 0.275 & 0.245 & 0.135 & 0.196 & 7.66 \\

  \end{tabular}
\end{table*}

We now derive the sonic scale for each simulation, defined as the spatial scale at which the characteristic velocity difference, $\delta v(\ell)$, equals the sound speed, $c_s$. Figure~\ref{Fig:sonic_scale} shows, in colours, the velocity-difference profiles $\delta v(\ell)$  normalised to the sound speed ($c_s$) as a function of { lag} $\ell$ for all models. 
{ For scales $\ell \gtrsim 0.2$ pc}, the curves follow an approximate power law with an index of $\sim 1/2$, consistent with expectations for a compressible turbulent medium, while the flattening at large scales coincides with the energy–injection range of each simulation. { On scales $\ell \lesssim 0.2$~pc the curves steepen, presumably as a consequence of numerical dissipation.} The horizontal red dashed line marks the sonic transition ($\delta v / c_s = 1$), and the corresponding sonic scales for each model are indicated by grey vertical dashed lines.

As shown in this figure, the sonic scale varies systematically with both the Mach number and the forcing scale. For a given forcing range, simulations with higher Mach number models exhibit smaller sonic scales, reflecting the stronger velocity differences induced by { the} stronger { driving}. Similarly, at a fixed Mach number, decreasing the forcing scale yields { larger velocity differences at a given wavenumber, and therefore} smaller sonic scales{ , because the same amount of injected energy is distributed over a narrower wavenumber range}. These trends highlight the sensitivity of the sonic scale to the underlying turbulent forcing, and call into question whether molecular clouds {, such as Lupus I, Polaris Flare, Taurus, Ophiuchus, and Orion,} have a sonic scale of the order of $\sim$0.1-0.27~pc \citep[][]{Ossenkopf_MacLow2002,Brunt2010, Yun+2021a, Yun+2021b, Ishihara2025}.

The main statistical results of the projected core separations are summarised in
Table~\ref{tab:core_stats}. The table lists the simulation label (first column), the physical forcing scale $\lambda_{\rm for} \equiv L_{\rm box}/k_{\rm for}$, (second column); the { derived} sonic scale $\ell_s$ (third column); the total number of detected cores $N_{\rm cores}$ (fourth column), the mean $\promedio{S}$ and median $S_{\rm med}$ projected nearest-neighbour separations (fifth and sixth columns, respectively), the standard deviation (seventh column), and the ratios $S_{\rm med}/\lambda_{\rm for}$ and $S_{\rm med}/\ell_s$, which quantify how the characteristic core spacing compares to turbulent driving and sonic scales. { From the latter column, it is clear that the ratio of the median core separation to the sonic scale depends on both the turbulent Mach number and the injection scale, implying that the core separation is not determined by the sonic scale.}

\begin{figure}
\includegraphics[width=0.45\textwidth]{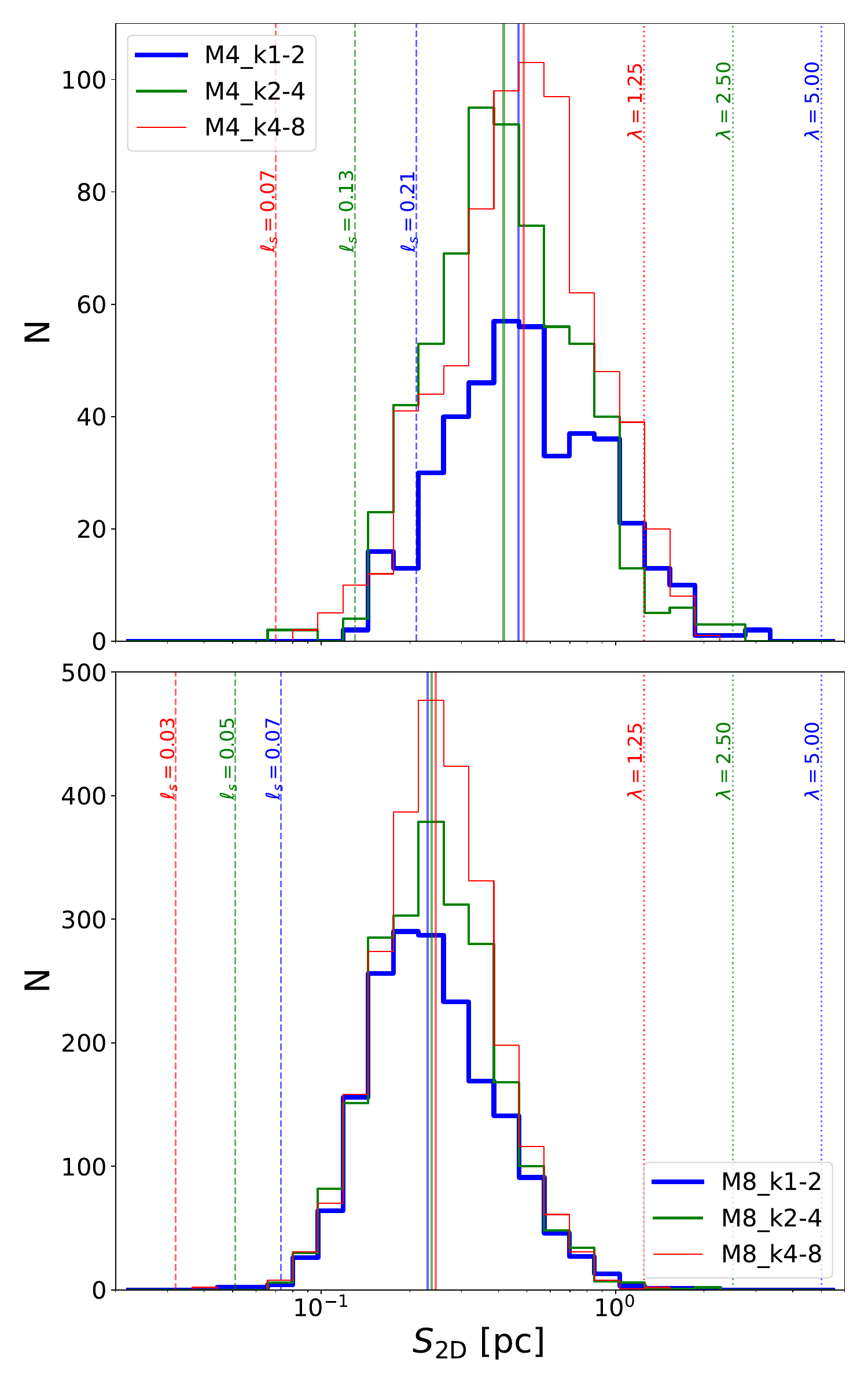}
\caption{Histograms of the nearest-neighbour distances ($S_{\rm 2D}$) with logarithmic binning for the simulations with $\mathcal{M} = 4$ (top panel) and $\mathcal{M} = 8$ (lower panel). Each histogram combines the $S_{\rm 2D}$ distributions from the three orthogonal projections. The horizontal lines indicate the sonic scale (dashed lines), the median $S_{\rm 2D}$ (solid lines), and the physical forcing scale ($\lambda_{\rm for} = L_{\rm box}/k_{\rm for}$; dotted lines).}
\label{fig:2D_histo}
\end{figure}

Figure~\ref{fig:2D_histo} shows the histograms of the projected core separations ($S_\mathrm{2D}$) for simulations with $\mathcal{M} = 4$ (top panel) and $\mathcal{M} = 8$ (bottom panel). Each panel includes results for the three turbulent–forcing ranges ($k_{\rm for}=1$--2, in black; $k_{\rm for}=2$--4, in blue; and $k_{\rm for}$~$=$4--8, in red). For each run, we indicate the sonic scale $\ell_s$ (vertical dashed lines), as inferred from Fig.~\ref{Fig:sonic_scale}, and the minimum energy–injection scale, $\lambda = L_{\rm box}/k_{\rm for}$ (vertical dotted lines). Together with Table~\ref{tab:core_stats}, these histograms highlight several noteworthy trends: 

\begin{enumerate}

 \item The number of identified dense cores (fourth column in Table~\ref{tab:core_stats}) is systematically higher for the Mach~8 than for the Mach~4 runs.
 
 \item The number of dense cores also increases consistently as the turbulent forcing scale decreases.

 \item Cores in higher-Mach number simulations tend to exhibit smaller characteristic separations than cores in lower-Mach number runs (see the fifth column of Table~\ref{tab:core_stats}. 

 \item The width of the separation distributions (column 7 in Table~\ref{tab:core_stats}) { is narrower for larger} Mach number. In addition, although the trend is marginal, the distribution width appears to decrease as the physical forcing scale becomes larger. 
 
 \item In all cases, the typical core separation remains significantly larger than the sonic scale (compare columns 3 and 5 in Table~\ref{tab:core_stats}). 

 \item  Finally, our adopted normalisation ensures that the mean separations obtained for the $\mathcal{M}=8$ models are in good agreement with the observational values reported by \citetalias{Ishihara2025} for the Polaris Flare ($\langle S_{\rm 2D} \rangle \simeq 0.21$~pc), although they are roughly twice as large as those measured in Lupus~I ($\langle S_{\rm 2D} \rangle \simeq 0.1$~pc).

\end{enumerate}

\begin{figure}
    \centering    \includegraphics[width=0.49\textwidth]{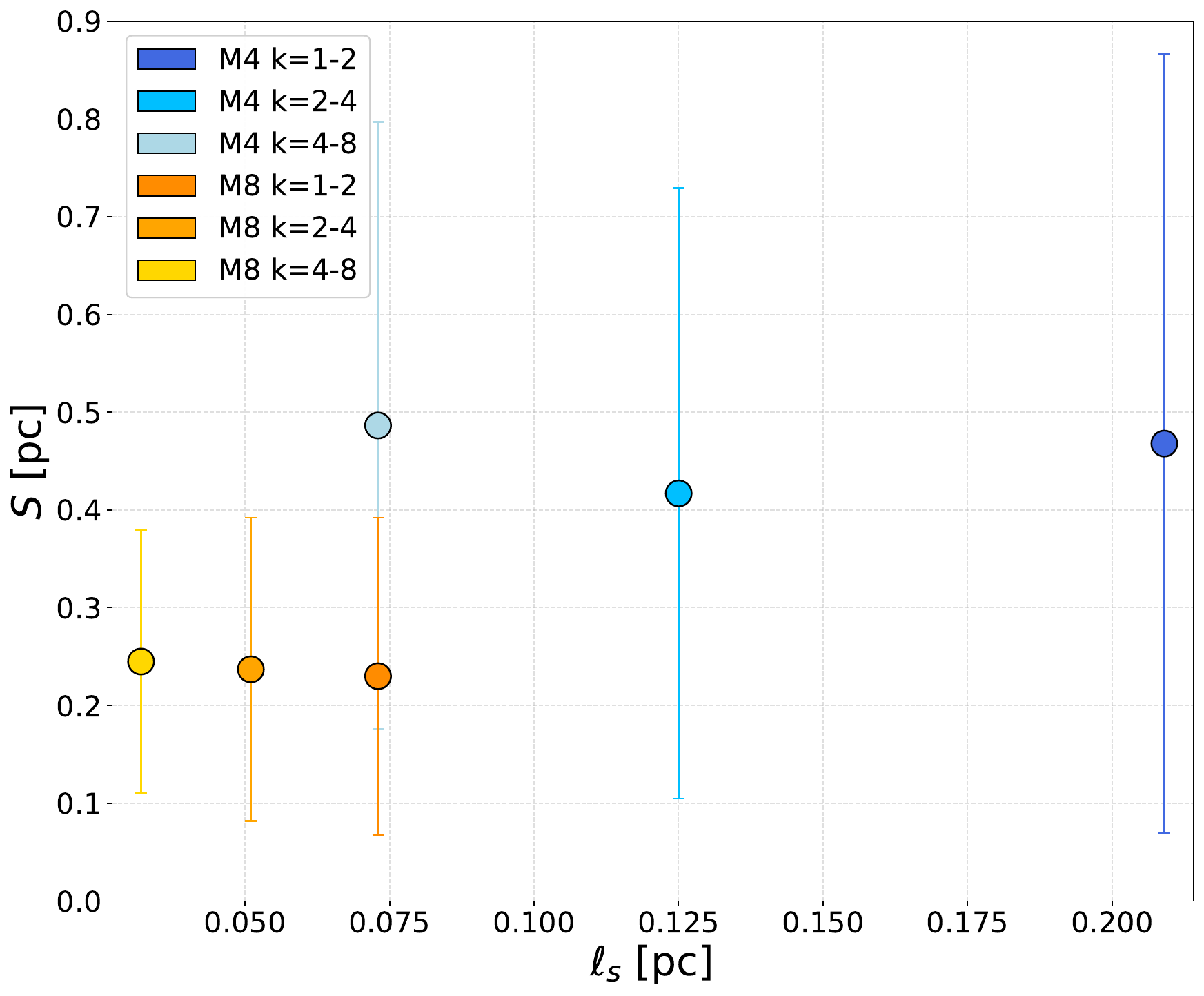}
    \caption{Median inter‑core separations versus sonic scale ($\ell_s$) (or forcing scale, $k_{\rm for}$) for the two turbulence regimes, $\mathcal{M} = 4$ (bluish colors) and $\mathcal{M} = 8$ (orangish colors). The error bars represent the standard deviation of each distribution. The systematic offset between boxes confirms that lower Mach numbers yield larger average separations at all forcing scales.}
    \label{fig:boxplot}
\end{figure}

In general, while the sonic scale varies systematically with both $\mathcal{M}$ and $k_{\rm for}$ (see Figure~\ref{Fig:sonic_scale}), the measured core separations exhibit no statistical correlation with $\ell_s$. This result is clearly illustrated in Figure~\ref{fig:boxplot}, which plots the median core separation as a function of the sonic scale for each simulation, with error bars indicating the dispersion of the distributions. No discernible trend is observed, reinforcing the conclusion that the core spacing is not governed by the sonic transition.

% /mnt/e/myProjects/Turb/python/plot_histo_M4_M8_lin.py

%%%%%%%%%%%%%%%%%%%%%%%%%%%%%%%%%%%%%%%%

\section{Discussion}

%Pensar en este par de parrafos:
% As discussed previously, the sonic scale ($\ell_s$) represents the characteristic length at which the turbulent velocity dispersion becomes comparable to the sound speed. This scale therefore marks the transition from supersonic motions that dominate at large scales to subsonic turbulence at smaller scales. Consequently, above $\ell_s$, the gas is expected to fragment through turbulence, whereas below this threshold gravitational fragmentation (described by the Jeans instability) becomes the dominant process.

% {\color{red} JBP comment: moví el siguiente párrafo de la sección discusión para acá, porque creo que no es cierta: lo que sistemáticamente vemos en las simulaciones es que conforme los cores se "condensan", también se acercan unos a otros... a menos de que se trate de una burbuja en expansión, como sugería Lee, en cuyo caso tampoco se preserva la distancia... }
% Assuming that the separation between cores is more likely a reflection of the initial sizes of the core precursors (fragments) produced by some physical mechanism. As these structures undergo individual contraction, their initial separation may largely be preserved. As the sonic scale is often smaller than the Jeans length, particularly in low-density regions (e.g., $\sim~3~\text{pc}$ in an MC with $<n>~=~100~\text{cm}^{-3}$), there is no physical reason for it to determine core separation.

Our results indicate that the median projected separation between dense cores, $S_{\rm 2D}$, depends on the Mach number, { decreasing} by approximately a factor of two when the Mach number is doubled. We interpret this behaviour as a direct consequence of { an enhanced} supersonic turbulent fragmentation as the Mach number increases.  In fact, at high turbulence levels ($\mathcal{M} = 8$), the typical core separation is relatively small, $\sim0.23~\mathrm{pc}$, while in the lower-Mach number runs ($\mathcal{M} = 4$), the separations are significantly larger, spanning $0.49$–$0.58~\mathrm{pc}$ (see Table~\ref{tab:core_stats} and Figure \ref{fig:boxplot}). This trend highlights the dominant role of the turbulent energy injection rate in regulating the characteristic spacing between dense structures. Moreover, the energy injection rate also controls the number of fragments formed{. Indeed, the} simulations with $\mathcal{M}=8$ produce approximately $\gtrsim 3$ times more cores than those with $\mathcal{M}=4$ (Table~\ref{tab:core_stats}, column 4), in agreement with previous studies \citep[e.g.][]{Girichidis+2011}.

Interestingly, although our results show a clear tendency for the number of cores to increase with the forcing wavenumber (Table~\ref{tab:core_stats}, column 4), the typical separations between them, measured either through the average or the median separation (columns 5 and 6) do not exhibit a systematic trend with the wavenumber. { This is most likely a consequence of the Mach number being larger in the excited scales below the injection scale when the driving is at smaller scales, since, as explained above, the same total energy is distributed over a narrower range of wavenumbers in this case (when the dissipation scale is kept fixed)}. On the other hand, there is a clear anti-correlation between the sonic scale and the number of cores (Table~\ref{tab:core_stats}, columns 3 and 4), the origin of which may be coincidental but requires a further study. { Also}, an interesting trend with the forcing scale comes from the width of the distributions (Table~\ref{tab:core_stats}, column 7). As the wavenumber increases in both Mach 4 and 8 simulations, the distribution becomes slightly narrower. 
 
We find no compelling evidence that core separations are preferentially set by the sonic scale. Instead, the characteristic spacing appears to be determined by the rate of turbulent energy injection, emphasising that fragmentation in turbulence-dominated clouds is externally regulated rather than emerging as an intrinsic feature of the turbulent cascade itself.

Our findings contrast with the interpretation proposed by \citetalias{Ishihara2025}, who argued that the observed agreement between core separations and the sonic scale in molecular clouds supports a universal turbulent fragmentation scenario. Although their observational measurements, such as those in the Polaris Flare and Lupus I clouds, indeed show separations comparable to the value of the sonic scale (typically assumed to be 0.1~pc), our simulations demonstrate that such agreement may occur coincidentally under certain forcing conditions rather than as a direct causal outcome of turbulence physics.

This distinction may have important implications for theoretical models of star formation. Our results suggest that the fragmentation scale in turbulence-dominated clouds is not uniquely determined by the sonic transition, { and} instead depends sensitively on the global properties of turbulent forcing.

In general, these findings support a more nuanced picture of turbulent fragmentation, in which the characteristic fragmentation length scale is not determined solely by internal turbulent properties (such as $\ell_s$), but also by how the turbulence is injected and sustained within the cloud. Consequently, interpreting the clustering of dense cores in observed clouds requires careful consideration of the underlying energy-injection mechanisms, which may vary substantially from cloud to cloud.

\subsection{Limitations}

In our suite of simulations, the sonic scale is not fully resolved in the models with the higher Mach number ($\mathcal{M} = 8$), primarily because the scale at which the gas transitions from supersonic to subsonic motions lies within the dissipative range of the velocity power spectrum. However, the simulations $\mathcal{M} = 8$ exhibit trends in characteristic velocity differences at a given scale, $\delta v (\ell)$, which are fully consistent with those obtained in the best resolved model, M4\_k1-2, where the sonic scale is well resolved (with 10 cells). This consistency suggests that our findings are not dominated by resolution effects.

We emphasise that resolving the sonic scale, especially in high-Mach number regimes, would require substantially higher computational cost \citep[e.g.,][]{Federrath+2021}. Therefore, our strategy of cross-validating trends across Mach numbers, anchored by well-resolved low-Mach simulations, provides a computationally efficient but physically reliable approach to understanding the role of the sonic scale in turbulent fragmentation.

\section{Summary and Conclusions}\label{sec:conclusions}

We have performed a suite of three-dimensional turbulent numerical simulations to investigate the physical origin of core separations under different turbulent regimes. Using a uniform grid of $512^3$ cells and varying both the Mach number and the turbulent forcing scale, we explored how these parameters influence the fragmentation process and the spatial distribution of dense cores. Our main results are summarised below:

\begin{itemize}
    \item The median projected core separation, $S_{\rm 2D,med}$, decreases with increasing Mach number, { being} approximately $\sim0.49-0.58$~pc for $\mathcal{M}=4$ runs and $\sim0.23-0.24$~ pc for $\mathcal{M}=8$ runs. The latter values are consistent with observational estimates for nearby low-density clouds such as the Polaris Flare and Lupus I \citepalias{Ishihara2025}. These separations are robust across different projections and remain stable throughout the simulations.

    \item No statistical correlation is found between the median core separation and the sonic scale, $\ell_s$. The core spacing does not trace the transition from supersonic to subsonic turbulence, indicating that the sonic scale does not determine the fragmentation length in our models.

    \item Instead, the core separation exhibits a clear and systematic dependence on the turbulent energy injection rate, controlled by the Mach number. Stronger { driving} produces denser clustering of cores, { suggesting} that global driving conditions, rather than intrinsic turbulent scales, set the characteristic spacing.

    \item These results challenge the common assumption that the sonic scale acts as a universal fragmentation threshold. The apparent observational agreement between $S_{\rm 2D}$ and $\ell_s$, as reported in some clouds, may therefore be coincidental and dependent on specific driving conditions, rather than a fundamental property of the turbulent cascade.
    
\end{itemize}

In addition to the points listed above, a relevant issue should be emphasised: in the interstellar medium, different clouds exhibit varying levels of turbulence, and thus the sonic scale is not necessarily the same for all molecular clouds in the Solar Neighbourhood (and certainly not the same through the Galaxy). Thus, the sonic scale is a scale that requires careful determination for each cloud, and it should not be assumed that some { well studied} clouds { (such as Polaris Flare, Lupus I, and other in the solar neighbourhood)}  exhibit a sonic scale { of the order of $\sim 0.1-0.27$ pc} { (see \citetalias{Ishihara2025} and references therein)}.

We conclude that in turbulent environments, the characteristic spacing of dense cores is not an intrinsic property of the turbulence itself, but rather a consequence of how the turbulence is driven. This insight has significant implications for both theoretical models and the interpretation of observations.

\section*{Acknowledgements}

MZA acknowledges support from SECIHTI grant number 320772. AP { and EVS} acknowledge financial support from the UNAM-PAPIIT IG100223 grant. { AP also acknowledges support from } the { SNII} of SECIHTI, M\'exico.
GCG acknowledges support from UNAM-PAPIIT grant number IN110824.
The authors thankfully acknowledge computer resources, technical advice and support provided by LANCAD-UNAM-DGTIC-442 and SECIHTI, through the use of the Miztli supercomputer at DGTIC–UNAM.

%%%%%%%%%%%%%%%%%%%%%%%%%%%%%%%%%%%%%%%%%%%%%%%%%%
\section*{Data Availability}

The data underlying this article will be shared on reasonable request to the corresponding author.

%%%%%%%%%%%%%%%%%%%% REFERENCES %%%%%%%%%%%%%%%%%%

% The best way to enter references is to use BibTeX:

\bibliographystyle{mnras}
\bibliography{refs} % if your bibtex file is called example.bib

%%%%%%%%%%%%%%%%%%%%%%%%%%%%%%%%%%%%%%%%%%%%%%%%%%

%%%%%%%%%%%%%%%%% APPENDICES %%%%%%%%%%%%%%%%%%%%%
%%%%%%%%%%%%%%%%% APPENDICES %%%%%%%%%%%%%%%%%%%%%
%
\appendix
%
%\section{Some extra material}
%something
%
\section{Dependence of astrodendro parameters} 
\label{app:astrodendro}

{
The \texttt{astrodendro} algorithm relies on three key parameters: \texttt{min\_value}, \texttt{min\_delta}, and \texttt{min\_npix}.  As in \citetalias{Ishihara2025}, we examined the influence of the first two parameters on core properties by varying \texttt{min\_value} in the range 1.16 to 4.16 and \texttt{min\_delta} in the range 0.54 to 2.08 in the model M8\_k1-2, using the $x$ projection at $t=4\times t_{\rm c}$. The results are summarised in Table~\ref{tab:dendro_params}. Note that the median of the column density is $N_0~=~2.16~\times~10^{21}~$cm$^{-2}$ and the taken fiducial parameters are \texttt{min\_value}$=0.95 \, N_0$ and \texttt{min\_delta}$= 0.475 N_0/2$ (grey row in the table).}

{
As shown in Table~\ref{tab:dendro_params}, varying the \texttt{astrodendro} parameters \texttt{min\_value} and \texttt{min\_delta} leads to significant changes in both the number of identified cores ($N_{\rm cores}$) and the median projected core separation ($S_{\rm 2D,med}$). For instance, increasing \texttt{min\_value} results in a decrease in the number of cores and a reduction in the median core separation, with values dropping from 243 cores and 0.241~pc for \texttt{min\_value} = 1.05 to 109 cores and 0.181~pc for \texttt{min\_value} = 4.05. Conversely, reducing \texttt{min\_delta} tends to increase the number of identified cores, as seen when lowering it from 2.06 to 0.515, leading to an increase in $N_{\rm cores}$ to 319, while the median core separation decreases slightly to 0.185~pc.

In the end, we adopt the fiducial parameters \citep[as in ][]{Chen+2018} as representative values for all our simulated maps, while noting that the number of cores and their separation are sensitive to the choice of dendrogram parameters.
}

\begin{table}
  \centering
  \caption{Dependence of core properties on \texttt{astrodendro} parameters.}
  \label{tab:dendro_params}
  \begin{tabular}{l r r}
    \hline
    (\texttt{min\_value}, \texttt{min\_delta})$\times 10^{21}$ & $N_{\rm cores}$ &
    $S_{\rm 2D,med}$ [pc] \\

    \hline
     (1.05, 1.03)   &   243    & 0.241  \\    
     (2.05, 1.03)   &  218  &  0.224  \\
     (3.05, 1.03)   &   144    &  0.196 \\
     (4.05, 1.03)   &   109    & 0.181  \\
     (2.05, 0.515)   &  319    & 0.185  \\
     (2.05, 2.06)   &   116    & 0.254  \\
    \hline
  \end{tabular}
\end{table}

% Don't change these lines
\bsp	% typesetting comment
\label{lastpage}
\end{document}